\documentclass[12pt,nofootinbib]{article}
\pdfoutput=1
\usepackage{slashed}
\usepackage{amsmath,amssymb,graphicx,multicol} 
\usepackage{epsf}
\usepackage[nosort]{cite}
\usepackage{cancel}
\usepackage{framed}
\usepackage{multirow}
\usepackage[usenames,dvipsnames]{color}

\usepackage{hyperref} 
\hypersetup{linktocpage=true}
\usepackage[all]{hypcap}
\hypersetup{
    bookmarks=true,         
    unicode=false,          
    pdftoolbar=true,        
    pdfmenubar=true,        
    pdffitwindow=false,     
    pdfstartview={FitH},    
    pdftitle={Taming the off-shell Higgs boson},    
    pdfauthor={Aleksandr Azatov, Christophe Grojean, Ayan Paul and Ennio Salvioni},     
    pdfsubject={hep-ph},   
    pdfcreator={TeXShop},   
    pdfproducer={MacTex, PDFLaTeX}, 
    pdfkeywords={Higgs Decays} {Effective Higgs Couplings} {Higgs to four leptons}, 
    pdfnewwindow=true,      
    colorlinks=true,       
    linkcolor=blue,          
    citecolor=magenta,        
    filecolor=magenta,      
    urlcolor=cyan           
}

\newcommand{\hhref}[1]{\href{http://arxiv.org/abs/#1}{arXiv:#1}}

\newcommand{\beq}{\begin{eqnarray}}
\newcommand{\eeq}{\end{eqnarray}}

\newcommand{\centeron}[2]{{\setbox0=\hbox{#1}\setbox1=\hbox{#2}\ifdim

\wd1>\wd0\kern.5\wd1\kern-.5\wd0\fi \copy0

\kern-.5\wd0\kern-.5\wd1\copy1\ifdim\wd0>\wd1
                                       \kern.5\wd0\kern-.5\wd1\fi}}
\newcommand{\ltap}{\>\centeron{\raise.35ex\hbox{$<$}}
                               {\lower.65ex\hbox{$\sim$}}\>}
\newcommand{\gtap}{\>\centeron{\raise.35ex\hbox{$>$}}
                               {\lower.65ex\hbox{$\sim$}}\>}

\newcommand\ZZ{\hbox{\zfont Z\kern-.4emZ}}
\font\zfont = cmss10 

\def\beq{\begin{equation}}
\def\eeq{\end{equation}}
\def\bea{\begin{eqnarray}}
\def\eea{\end{eqnarray}}

\def\bit{\begin{itemize}}
\def\eit{\end{itemize}}

\def\l{\left}
\def\r{\right}

\def\ra{\rightarrow}

\def\baa{\begin{array}}
\def\eaa{\end{array}}

\def\d{\partial}

\def\simgt{\mathrel{\lower2.5pt\vbox{\lineskip=0pt\baselineskip=0pt
           \hbox{$>$}\hbox{$\sim$}}}}
\def\simlt{\mathrel{\lower2.5pt\vbox{\lineskip=0pt\baselineskip=0pt
           \hbox{$<$}\hbox{$\sim$}}}}

\def \dblarrow#1{\stackrel{\longleftrightarrow}{#1}}
\def\cM {{\cal M}}

\begin{document}
\begin{titlepage}
\begin{flushright}
{\tt CERN-PH-TH-2014-108}
\end{flushright}


\begin{center}
{\Large \bf  
Taming the off-shell Higgs boson 
}
\end{center}
\vskip0.5cm

\renewcommand{\thefootnote}{\fnsymbol{footnote}}
\begin{center}
{\large Aleksandr Azatov$^{a}$, Christophe Grojean$^{b}$, Ayan Paul$^{c}$ and Ennio Salvioni$^{d}$
\footnote{email:  Aleksandr.Azatov@cern.ch, Christophe.Grojean@cern.ch, Ayan.Paul@roma1.infn.it and esalvioni@ucdavis.edu.}
}
\end{center}

\begin{center}
\centerline{$^{a}${\small \it Theory Division, Physics Department, CERN, CH-1211 Geneva 23, Switzerland}}
\vskip 4pt
\centerline{$^{b}$ {\small \it ICREA at IFAE, Universitat Aut\`onoma de Barcelona, E-08193 Bellaterra, Spain}}
\vskip 4pt
\centerline{$^{c}$ {\small \it INFN, Sezione di Roma, I-00185 Rome, Italy}}
\vskip 4pt
\centerline{$^{d}$ {\small \it Physics Department, University of California, Davis, CA 95616, USA}}
\end{center}

\renewcommand{\thefootnote}{\arabic{footnote}}

\vglue 1.0truecm

\begin{abstract}
\noindent We study the off-shell Higgs data in the process $pp\to h^{(*)} \to Z^{(\ast)}Z^{(\ast)}\to 4\ell$, to constrain deviations of the Higgs couplings. We point out that this channel can be used to resolve the long- and short-distance contributions to Higgs production by gluon fusion and can thus be complementary to $pp\to ht\bar t$ in measuring the top Yukawa coupling. Our analysis, performed in the context of Effective Field Theory, shows that current data do not allow one to draw any model-independent conclusions. We study the prospects at future hadron colliders, including the high-luminosity LHC and accelerators with higher energy, up to 100 TeV. The available QCD calculations and the theoretical uncertainties affecting our analysis are also briefly discussed.

\end{abstract}

\end{titlepage}

\section{Introduction}
 
With the discovery of the Higgs boson by the ATLAS and CMS experiments~\cite{ATLASdiscovery,CMSdiscovery}, high energy physics experiences a transition: after a long period of search and exploration, an era of consolidation and precise measurements has just started and it will complement the direct search for new physics beyond the Standard Model (SM). With a mass around 125\,GeV, the Higgs boson offers various production modes and decay channels directly accessible to observation, supplying a wealth of data that can be used to learn about the Higgs couplings. In the absence of any indication of new light degrees of freedom below the TeV scale, the effects of BSM physics can be conveniently parameterized in terms of higher dimensional operators for the SM fields. This Effective Field Theory (EFT) approach relates Higgs data to measurements of other sectors of the SM, like ElectroWeak (EW) precision data, and it gives a systematic way for controling the deviations away from the SM, organized as an expansion in powers of the ratio of the momentum over the new physics scale. So far a lot of information has been extracted from inclusive rates, which are dominated by resonant production of the Higgs boson near the mass peak, i.e. at scales close to the Higgs mass itself.  
 
As for any other quantum particle, the influence of the Higgs boson  is not limited to its mass shell. Recently, the CMS and ATLAS collaborations reported the differential cross-section measurement of $pp \to Z^{(*)}Z^{(*)} \to 4\ell,2\ell2\nu$ ($\ell = e, \mu$) at high invariant-mass of the $ZZ$ system~\cite{CMS,Khachatryan:2014iha,ATLASoffshell}. This process receives a sizable contribution from a Higgs produced off-shell by gluon fusion~\cite{glover,Kauer:2012hd}. 
As such, this process potentially carries information relevant for probing the EFT at large momenta and could thus reveal the energy-dependence of the Higgs couplings controlled by higher-dimensional operators with extra derivatives. It has been proposed~\cite{melnikov} to use the off-shell Higgs data to bound, in a model-independent way,  the Higgs width. However, as it was emphasized in Ref.~\cite{Englert:2014aca}, this bound actually holds under the assumption that the Higgs couplings remain the same over a large range of energy scales. The EFT Lagrangian captures and organizes precisely this energy-dependence of the Higgs couplings and therefore offers a coherent framework for analyzing the off-shell Higgs data. 
The situation seems a priori similar to the precision measurements of the EW observables, where off-shell $Z$ data at LEP2 complemented the $Z$-peak data and bounded $\mathcal{O}(p^4)$ dimension-6 operators, like the $W$ and $Y$ oblique parameters~\cite{Barbieri:2004qk}, in addition to the $\mathcal{O}(p^2)$ dimension-6 operators, the $S$ and $T$ oblique parameters~\cite{Peskin:1990zt}, already probed at LEP1.  However,  a careful exploration of the complete list of all dimension-6 operators deforming the SM Lagrangian\footnote[1]{Throughout this paper, we shall be working under the assumption that the Higgs boson is a part of an EW doublet. This assumption was not made in Ref.~\cite{Gainer:2014hha}, where the off-shell Higgs data was used to bound deviations of the Higgs couplings that, in the doublet realization, are either sub-dominant or correlated with other data from better measurements.} reveals~\cite{Elias-Miro:2013mua,Pomarol:2013zra} that the operators modifying the Lorentz structure of the SM Higgs couplings are already severely constrained by EW precision data or by the bounds on anomalous gauge couplings. Thus, qualitatively, the off-shell data do not open a new window, i.e. they do not probe new dimension-6 operators. 

Quantitatively, it is still worth exploring the actual bounds set by the off-shell data. Out of the eight $CP$-even dimension-6 operators uniquely probed by Higgs physics, five are already bounded by the decay channels of an on-shell Higgs boson. While double Higgs production, which could apprise us of the Higgs self-interaction, will mostly remain out-of-reach at the LHC, the two additional channels, $h \to Z\gamma$ and $pp \to t\bar t h$, will soon be accessible at  run 2 of the LHC operation~\cite{Dawson:2013bba} and should bound two extra dimension-6 operators that remain unconstrained as yet. 
The latter channel will be particularly important to unambiguously pin down the top Yukawa coupling which, at the moment, is accessed only radiatively via its one-loop contributions to the $gg \to h$ and $h \to \gamma \gamma$ processes. It is well known that these two processes alone cannot resolve the top loop and distinguish it from effective contact interactions of the Higgs boson to gluons or photons, which parameterize the effect of possible new physics at short distances. Therein lies the importance of the $t\bar t h$ channel.\footnote{It has recently been pointed out that the measurement of the ratio $\sigma(t\bar t h)/\sigma(t\bar t Z)$ at very high energy could provide a very clean access to the top Yukawa coupling \cite{MLM-Frederix}. We also recall that the top Yukawa coupling can be constrained indirectly by the study of top pair production near threshold at future $e^+ e^-$ colliders (see for instance Ref.~\cite{Martinez:2002st}).} However, an accurate measurement of this process is known to be challenging, due to its suppressed cross section and to the high multiplicity of its final states. The latter implies that obtaining accurate predictions for some of the relevant backgrounds, such as for example $pp \to t\bar{t}b\bar{b}$ for the $h \to b\bar{b}$ channel, is a difficult task. Alternative and complementary ways to separate the long- and short-distance contributions to the $ggh$ vertex are therefore welcome. Recently, it was proposed~\cite{Banfi:2013yoa,Azatov:2013xha,Grojean:2013nya,Schlaffer:2014osa} to study the hard recoil of the Higgs boson against an extra jet \cite{Ellis:1987xu,Baur:1989cm,Abdullin:1998er}, which provides a second scale above the Higgs mass to probe the EFT structure (see also Ref.~\cite{Buschmann:2014twa} for a study of $h + 2\,\mathrm{jets}$). The double Higgs production by fusion of gluons also effectively introduces a second mass scale and can be used to separate the top Yukawa coupling from the contact interaction to gluons or photons~\cite{Contino:2012xk,Gillioz:2012se}. Note that these two channels will require some large integrated luminosity, beyond the run 2 of the LHC. In this paper we want to advocate that off-shell Higgs production is another obvious place to break this degeneracy of the couplings and to learn about the top Yukawa coupling.

One advantage of the analysis of Higgs data in terms of an EFT, over a simple fit in terms of anomalous couplings, is that it comes with some simple consistency checks that guarantee the reliability of its results against our ignorance of the details of the new physics sector. For instance, it is possible to say when it is safe to neglect dimension-8 operators over the dimension-6 ones. As we are going to illustrate, this is of prime importance when the data is not strong enough to derive stringent bounds. In particular, we shall see that no model-independent reliable bounds can be extracted from the 8\,TeV data. The situation improves at 14\,TeV and upon accumulating a luminosity of about 3\,ab${}^{-1}$ it will be possible to derive meaningful constraints, at least for rather strongly coupled new physics. Only at future, higher-energy accelerators, however, do the bounds become truly model-independent.

This paper is organized as follows. In Section~\ref{sec:prod}, we present our analysis of the Higgs couplings using the 8\,TeV off-shell data and we discuss the reliability of the results in an EFT framework. In Section~\ref{sec:HighE}, we study the prospects of the off-shell study at future facilities like the high-luminosity LHC and very high-energy hadron-hadron colliders. We conclude in Section~\ref{sec:conclusion}, whereas some technical details are collected in three Appendices.

\section{Constraining the anomalous couplings of the Higgs boson}
\label{sec:prod}

Recently a new method was suggested in Ref.~\cite{melnikov} to indirectly constrain the Higgs width, by looking at the very high invariant mass region of the four-lepton final state in the $pp \to Z^{(*)}Z^{(*)}\to 4\ell$ channel, which receives contributions from the exchange of a highly off-shell Higgs, and comparing the event yields with the SM predictions. More precisely, information on the Higgs width can be extracted by comparing the event yields off and on the Breit--Wigner peak. It follows that this method relies on the following assumptions:
\begin{itemize}
\item there is an invisible Higgs decay width, so that the total width of the Higgs and its couplings can be varied independently;
\item variations of all the Higgs couplings are universal;
\item there are no higher dimensional operators affecting either Higgs decay or production.
\end{itemize}
In this paper we will use the same process $pp \to Z^{(*)}Z^{(*)}\to  4\ell$ to put constraints on new physics, however we will reverse the first and third assumptions: we will assume the absence of an invisible decay width and the presence of new higher dimensional operators which can modify the production or decay of the Higgs boson. The second assumption stated above was necessary in Ref.~\cite{melnikov}, to keep the 
Higgs on-shell measurements SM-like. In our analysis we will not make this assumption, however we will make sure that the parameter space we explore is not excluded by the on-shell Higgs measurements.

\subsection{Operators contributing to Higgs production}
\label{sec:prod_ops}
Let us start by considering the operators affecting Higgs production by gluon fusion. Assuming the Higgs boson to be part of a doublet of $SU(2)_{L}$, there are three relevant dimension-6 operators\footnote{The operator $O_H=\l(\d(H^\dagger H)\r)^2$ also leads to a modification of the top Yukawa coupling and thus affects the Higgs production by gluon fusion. However, the constraints  on its Wilson coefficient $c_{H}$ obtained by combining information from the various on-shell Higgs channels are generically much stronger than those on $c_y, c_g$, so we will ignore the effects of $O_{H}$ in this paper. Also, at the dimension-6 level, there are dipole operators which can modify both the signal and the background:
\bea \label{dipoles}
\bar Q_L \widetilde{H} \sigma_{\mu\nu}t_R B_{\mu\nu}+\mathrm{h.c.}\,,\qquad \bar Q_L \sigma^a \widetilde{H} \sigma_{\mu\nu}t_R W^a_{\mu\nu}+\mathrm{h.c.}\,,\qquad \bar Q_L \widetilde{H} \sigma_{\mu\nu}t_R G_{\mu\nu}+\mathrm{h.c.}\,.
\eea
However, their effects usually have an additional loop-suppression compared to those of $c_{y},c_{g}$ and anyway these operators will be better constrained by top data alone. Therefore these dipole operators will also be neglected in our analysis.
}
\bea
\label{d6}
&& {\cal L}^{\hbox{\small dim-6}} = c_y \frac{y_t |H|^2}{v^2} \bar{Q}_L \widetilde{H} t_R+\mathrm{h.c.}+\frac{c_gg_s^2}{48\pi^2 v^2}|H|^2G_{\mu\nu}G^{\mu\nu}+\tilde c_g \frac{g_s^2}{32\pi^2 v^2} |H|^2 G_{\mu\nu} \tilde G^{\mu\nu}\,,\nonumber\\
&&\ \ \  \hbox{with} \ \ \tilde G_{\mu\nu}=\frac{1}{2}\epsilon^{\mu\nu\lambda\rho}G_{\lambda\rho}\,,
\eea
where $v\simeq 246\,\mathrm{GeV}$ is the Higgs vacuum expectation value. Notice that, given our normalization, the parameterization of new physics effects in terms of an EFT expansion is meaningful only if the Wilson coefficients satisfy 
\begin{equation}
c_{i}\ll 1\, .
\end{equation}
After electroweak symmetry breaking Eq.~\eqref{d6} leads to the Lagrangian
\bea
\label{nonl}
{\cal L}&=&-c_t \frac{m_t}{v}\bar t t h+\frac{g_s^2}{48 \pi^2} c_g \frac{h}{v}G_{\mu\nu}G^{\mu\nu},
\eea
where $c_t=1-\mathrm{Re}(c_y)$ and we have ignored $CP$-odd contributions. It is well known (see for instance Refs.~\cite{Azatov:2013xha, Grojean:2013nya}) that the current measurements of the Higgs couplings have a strongly degenerate solution along the line $c_t+c_g= \mathrm{constant}$, which originates from the Higgs low-energy theorem: because on-shell Higgs production occurs at the scale $m_h<m_t$, its cross section is proportional to 
\bea
\label{xseconpeak}
\sigma\sim |c_t+c_g|^2\,.
\eea
However, once we go to the far off-shell region, the partonic center-of mass energy of the process $\sqrt{\hat{s}}$ becomes higher than $m_{t}\,$, so that we cannot integrate out the top anymore and Eq.~\eqref{xseconpeak} becomes invalid. Therefore comparing the measurements of the on-shell and off-shell Higgs production provides a way to disentangle the effects of the $c_t,c_g$ couplings. 
\begin{figure}
\begin{center}
\includegraphics[scale=0.7]{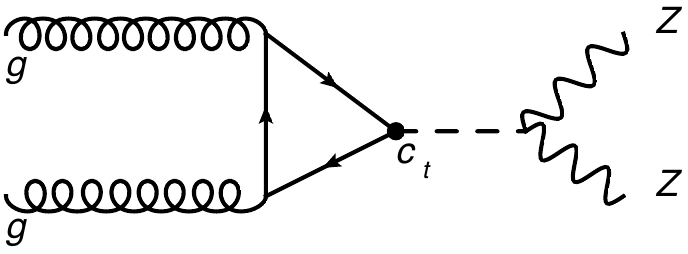}\hspace{0.6cm}
\includegraphics[scale=0.7]{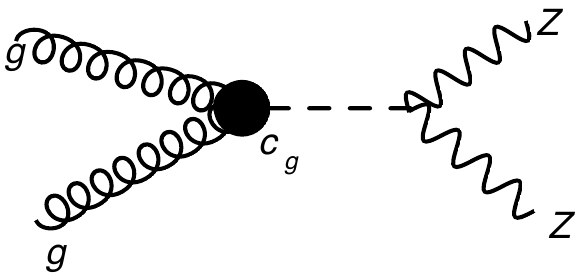}\hspace{0.6cm}
\includegraphics[scale=0.7]{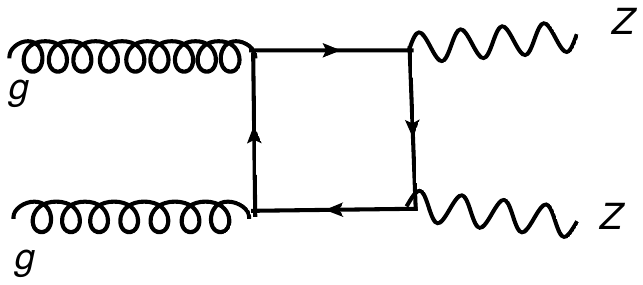}
\caption{Sample diagrams contributing to $gg\ra ZZ$. 
\label{diag}}
\end{center}
\end{figure}
Fig.~\ref{diag} shows the diagrams contributing to the $gg\ra ZZ$ process, whose amplitude can be schematically written as
\bea \label{amp}
\mathcal{M}_{gg\to ZZ} = \mathcal{M}_{h} + \mathcal{M}_{bkg}= c_{t}\mathcal{M}_{c_{t}}+c_{g}\mathcal{M}_{c_{g}}+\mathcal{M}_{bkg}\,,
\eea
where $\cM_h$ stands for the Higgs-mediated part, and $\cM_{bkg}$ stands for the interfering background, given by the box diagrams in Fig.~\ref{diag}.
Notice that in addition to the interfering $gg\ra ZZ$ background there is also a non-interfering irreducible background, produced by the $q\bar{q}\ra ZZ$ process. The SM amplitude for $gg\ra ZZ$ was computed for the first time in Ref.~\cite{glover}. As pointed out in Ref.~\cite{Kauer:2012hd}, the off-shell Higgs contribution is enhanced for on-shell $Z$ bosons, which makes the $\sqrt{\hat{s}}\gg 2m_{Z}$ region particularly relevant for Higgs couplings measurements. It is interesting to observe that the amplitude generated by the $c_g$ coupling grows with partonic center-of-mass energy $\sqrt{\hat s}$ like
\bea
\mathcal{M}^{++00}_{c_g}\sim \hat s\,,
\eea
to be compared to the triangle amplitude mediated by the top loop, which grows like
\bea
\label{Msm}
\mathcal{M}^{++00}_{c_t}\sim \log^{2} \frac{\hat s}{m_{t}^{2}},
\eea
in the notation for helicity amplitudes of Ref.~\cite{glover}.\footnote{In the SM, in the large $\sqrt{\hat s}$ region there is a strong cancellation between the triangle and the box contributions to the $gg\ra ZZ$ process~\cite{glover,Accomando:2007xc}. One can understand its origin by performing an $s$-channel cut of the loops and looking at the perturbative unitarity preservation in the $t\bar{t}\to ZZ$ subprocess. Note that for couplings different from those of the SM there is also unitarity violation directly in the $gg\ra ZZ$ process, due to the growth of the amplitude $\propto \log^2 \hat s$. However, this growth leads to a scale of unitarity violation that is exponentially high, $\Lambda \gtrsim  10^{13}$ GeV (computed requiring $\mathcal{M}\sim 16\pi$), and thus irrelevant for phenomenological purposes. We would like to thank R.~Contino for bringing these issues to our attention.}
%
%
  Thus for ${\hat s} \gg m_{t}^{2}$ the discriminating power of the off-shell Higgs production becomes stronger. However, at very high energies the EFT approximation breaks down and the dimension-8 operators become as important as the dimension-6 ones. For example, let us consider the operator
\bea
\label{d8}
O_8=\frac{c_8 g_s^2}{16\pi^2 v^4} G_{\mu\nu}G^{\mu\nu}\l(D_\lambda H\r)^\dagger D^\lambda H\,.
\eea
The matrix element corresponding to the final state with two longitudinally polarized $Z$ bosons grows with energy as
\bea
\mathcal{M}_{c_8}^{++00}\sim \hat{s}^2.
\eea
Then the interference of $O_{8}$ with the SM amplitude will become of the same order as the interference of the dimension-6 operators with the SM at the scale
\bea
\sqrt{\hat s}\sim \sqrt{\frac{c_g,c_y}{c_8}}\,v\,.
\eea
Therefore our analysis, based on Eq.~\eqref{d6}, is valid only up to this scale and it would not make sense to include the region with larger $\sqrt{\hat{s}}$. Furthermore, when squaring Eq.~\eqref{amp}, the terms in the cross section that are proportional to $c_{g,y}^2$  effectively behave like dimension-8 operators, as opposed to the terms linear in $c_{g,y}$ which constitute the true dimension-6 effects resulting from the interference with the SM amplitude. The contribution of $O_{8}$ is subleading with respect to the quadratic terms if
\bea \label{lvsnl}
c_{8}\ll c_{g,y}^{2}\,.
\eea
Whether this condition is satisfied or not, and thus, whether it is sensible to retain the quadratic terms or not, is a model-dependent question. Therefore, in the following we will present results for both cases: the `nonlinear' analysis, where the terms $\sim c_{g,y}^{2}$ are retained, and the `linear' analysis, where only the genuine dimension-6 effects are considered. The difference between the nonlinear and linear results becomes negligible for very small perturbations of the SM. However quantitatively we will find that, in the light of the current and future sensitivity of the off-shell Higgs measurement, this difference can be sizable. Finally, it is worthwhile mentioning that a significant difference between nonlinear and linear results does not arise for the $pp\to h + \mathrm{jet}$ process, which provides an independent handle on the $c_{t},c_{g}$ degeneracy.

\subsection{Bounds on  the Higgs couplings}

In order to find constraints on the Higgs couplings $c_t,c_g$ we need to know the differential cross section for $pp\to Z^{(\ast)}Z^{(\ast)}\to 4\ell$, $d \sigma /d m_{4\ell}$, as a function of the four-lepton invariant mass $m_{4\ell}\equiv \sqrt{\hat{s}}$. The diagrams mediated by the Higgs boson exchange are functions only of $\sqrt{\hat{s}}$, therefore the differential cross section can be parameterized as 
\bea
\label{Fi}
\frac{d \sigma }{d m_{4\ell}}=F_0(m_{4\ell})+F_1(m_{4\ell}) c_R^2 +F_2(m_{4\ell}) c_I^2+F_3(m_{4\ell}) c_R+F_4(m_{4\ell}) c_I\,,
\eea
where $c_I$ and $c_R$ are defined as the ratios of the Higgs-mediated amplitudes compared to the SM values (the NP subscript stands for the new physics contribution)
\bea
c_R=\frac{\mathrm{Re}\, \cM_\Delta^{\hbox{\tiny NP+SM}} }{\mathrm{Re}\, \cM_\Delta^{\hbox{\tiny SM}} },~~c_I=\frac{\mathrm{Im}\, \cM_\Delta^{\hbox{\tiny NP+SM} }}{\mathrm{Im}\, \cM_\Delta^{\hbox{\tiny SM}}}\,,
\eea
where it is understood that $c_{R,I}$ depend also on $m_{4\ell}$. By varying the mass of the particle running in the triangle diagram,  we can easily extract the functions $F_{0, \ldots, 4}$ for any given $m_{4\ell}$.
 We modified the MCFM6.8 code~\cite{Campbell:2010ff,Campbell:2013una} in order to perform this procedure. Then the functions $F_i$ can be obtained from the following set of equations:
{\allowdisplaybreaks
\bea \label{method}
&&\hbox{Only signal:} \   | {\cal M}  _h|^2\sim F_1+F_2\,,\nonumber\\
&&\hbox{Only interference:}\ | {\cal M} _h+ {\cal M} _{bkg}|^2-| {\cal M} _h|^2-| {\cal M} _{bkg}|^2\sim F_3+F_4\,,\nonumber\\
&&\hbox{Only interfering background:}\ | {\cal M} _{bkg}|^2\sim F_0\,,\\
&&\hbox{Only signal with $m_t=M$:}\ | {\cal M} _h|_{m_t=M}^2\sim F_1 c_R(M)^2+F_2 c_I(M)^2\,,\nonumber\\
&&\hbox{Only interference with $m_t=M$:}\nonumber\\
&& \ \ \ \ | {\cal M} _{h(m_t=M)}+ {\cal M} _{bkg}|^2-| {\cal M} _{h(m_t=M)}|^2-| {\cal M} _{bkg}|^2\sim  F_3c_I(M)+ F_4 c_R(M)\,.\nonumber
\eea
}
We have checked that our method of extracting the functions $F_i$ is consistent by varying the input parameter $M$.
Then one can easily translate $(c_R,c_I)$ into the $(c_t,c_g)$ basis using the well-known expression for the triangle amplitude,
\bea \label{ds_ctcg}
\frac{d \sigma (c_t,c_g)}{d m_{4\ell}}=F_0+F_1 \l(c_t+c_g\frac{F_\Delta(\infty)}{\mathrm{Re}\, F_\Delta(m_t)} \r)^2 +F_3\l(c_t+c_g\frac{F_\Delta(\infty)}{\mathrm{Re}\,F_\Delta(m_t)} \r) +F_2 c_t^2+F_4 c_t\,,\nonumber\\
\eea
where $F_\Delta$ is the fermionic leading order loop function for single Higgs production (see Appendix~\ref{sec:loop} for the explicit  expression). We emphasize that this method of extracting coefficients works because the overall production cross section of the Higgs-mediated diagrams depends only on $\hat{s}$, without any dependence on the $\hat{t},\hat{u}$ variables. As we mentioned in Section~\ref{sec:prod_ops}, in the large invariant mass region there is a cancellation between the box and the triangle diagrams. 
This property of the amplitude leads to the following relations between the functions $F_i\,$, which we have verified numerically
\bea
\l.\frac{F_1+F_2}{F_0}\r|_{m_{4\ell}\ra \infty}=-\l.\frac{F_3+F_4}{2F_0}\r|_{m_{4\ell}\ra \infty}=\,1\,.
\eea 
To obtain the current bounds on the $(c_t,c_g)$ parameters we have used the results presented in Ref.~\cite{CMS}. In order to simplify our analysis we have decided to focus on the simple counting analysis, without using the results obtained with the application of the Matrix Element Likelihood Method (MELA) \cite{CMS,Khachatryan:2014iha}. The interested reader is referred to Appendix~\ref{cmsHwidth}, where the details of the analysis are presented. We would like to stress that we made use of MCFM only to compute the signal and the interfering background in $gg\ra ZZ$, whereas for the non-interfering background $q\bar{q}\ra ZZ$ the results presented by CMS were used.

\begin{figure}[t]
\begin{center}
\includegraphics[scale=0.7]{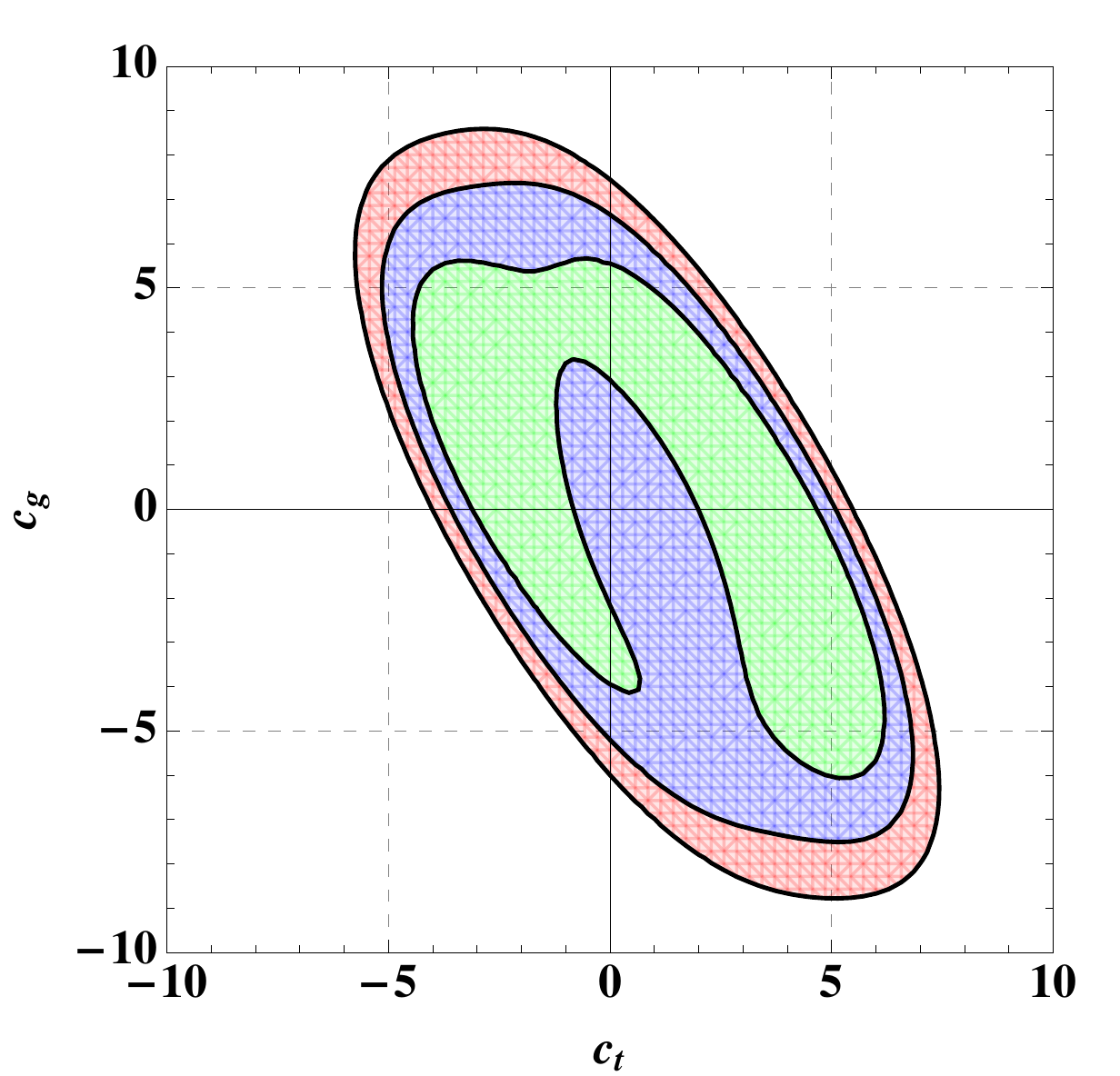}
\end{center}
\caption{\label{ctcg} $68\%,95\%$ and $99\%$ probability contours in the $c_t$,$c_g$ plane, using the 8\,TeV CMS data set. A 10\% systematic uncertainty was assumed on the $q\bar{q}$ background.}
\end{figure}

\begin{figure}
\begin{center}
\includegraphics[scale=0.7]{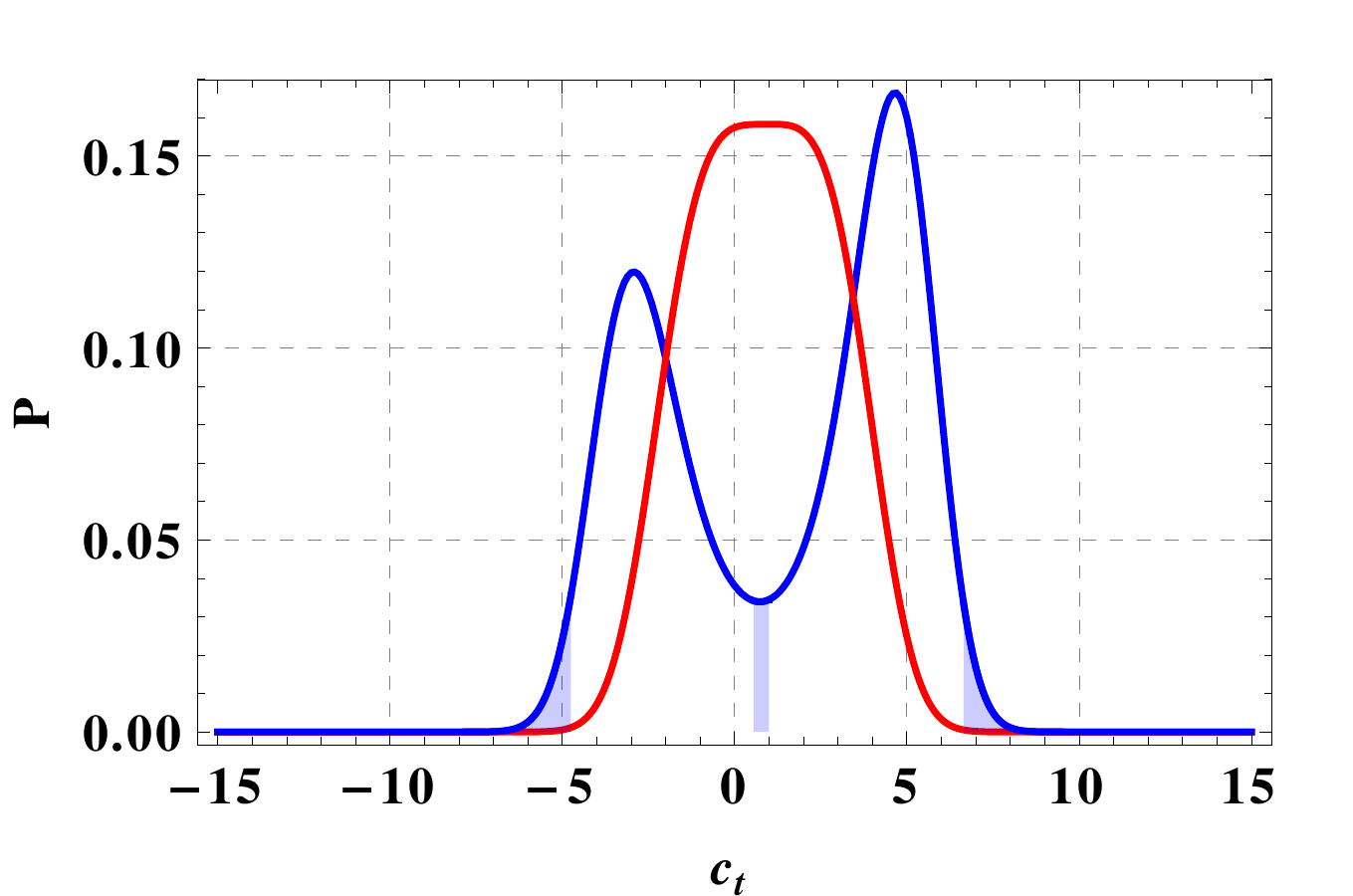}
\end{center}
\caption{\label{ct}Posterior probability as a function of $c_t$, assuming the constraint $c_t+c_g=1$, for the 8\,TeV CMS data set. At $95\%$ we find $c_t\in[-4.7,0.5]\cup[1, 6.7]$ (unshaded region), at $68\%$ $c_t\in [-4,-1.5]\cup[2.9,6.1]$. The red line shows the expected probability for the SM signal.}
\end{figure}

The resulting  constraints  in the  $(c_t,c_g)$ plane are shown in Fig.~\ref{ctcg}. 
In order to explore the power of resolving the $c_t$ vs.~$c_g$ degeneracy, we assume that the inclusive measurement is consistent with the SM and therefore we impose the condition $c_t+c_g=1$. The resulting posterior probability is presented in Fig.~\ref{ct}: with $68\%$ probability the coupling $c_t$ is constrained within $[-4,-1.5]\,\cup\,[2.9,6.1]$. These results were obtained using the nonlinear analysis. The CMS bound allows $c_{g,y}$ to be of $O(1)$, thus no interpretation of the results in terms of the EFT can be made. The bounds we quote here should therefore be understood as holding under the assumption that Eq.~\eqref{nonl} fully encodes the effects on $gg\to ZZ$ of the new physics, even though the latter is allowed to be at the weak scale. Finally, notice that our results were obtained using only the four-charged lepton final state and without the MELA, so upon a more refined analysis one can easily expect a factor of two improvement on the bounds on the couplings. 

Lastly, we wish to comment about higher-dimensional operators affecting the Higgs coupling to the $Z$ bosons, thus modifying the total number of events in $gg\ra h^{(*)} \ra ZZ$, which were studied in Ref.~\cite{Gainer:2014hha}. Assuming the Higgs to be part of an $SU(2)_{L}$ doublet, the operators whose contributions grow with energy more rapidly than that of the Standard Model appear only at the dimension-8 level, so the bounds on the scale of the new physics are weak (see Appendix~\ref{hdecay}). 

\section{Prospects at the High-Luminosity LHC and hadron-hadron Future Circular Colliders}
\label{sec:HighE}

In this section we turn our attention to the future of high-energy physics, and discuss the prospects of off-shell Higgs measurements at future proton-proton colliders. We consider the High-Luminosity LHC (HL-LHC), with a nominal energy and integrated luminosity of $14$\,TeV and $3$\,ab$^{-1}$ respectively, and the hadron-hadron Future Circular Colliders (FCC-hh), with energy varying from $33$ to $100$\,TeV. The physics case for the HL-LHC includes a program of high-precision Higgs couplings measurements, as well as the accessibility of new processes, such as double Higgs production, which could apprise us of the Higgs self-interaction. Exploration of the physics potential of the FCC-hh started only recently, and here we wish to contribute to that effort by performing a first estimate of the opportunities available in off-shell Higgs measurements.

\subsection{Details of the simulation}

The $gg\to ZZ$ process was simulated with MCFM6.8. To extract the cross section as a function of $c_t$ and $c_g$ we modified the code, in order to vary the top mass in the Higgs-mediated diagrams without modifying the $gg\to ZZ$ interfering background (see Eq.~\eqref{method}). It should be noted that MCFM also includes the loops of bottom quarks for the Higgs-mediated diagrams. However, since we did not consider modifications of the $b$-quark Yukawa couplings, these loops are effectively absorbed into the  interfering background in our parameterization of Eq.~\eqref{ds_ctcg}. The noninterfering, $q\bar{q}\to ZZ$ background was also generated using MCFM6.8.

%
 An important issue that must be taken into account when simulating $gg\to ZZ$ is that the Higgs contribution is known to Next-to-Leading Order (NLO, $O(\alpha_s^3)$) 
\cite{
Djouadi:1991tka,Graudenz:1992pv,
Spira:1995rr}
in QCD with exact top mass dependence and to Next-to-Next-to-Leading Order (NNLO,  $O(\alpha_s^4)$)\cite{Harlander:2002wh,Anastasiou:2002yz,Ravindran:2003um
} in the infinite top mass limit, whereas the interfering background is known only at leading order (LO,  $O(\alpha_s^2)$). As a consequence, assessing the higher order corrections to the full process is problematic, and several proposals have been put forward \cite{Passarino:2012ri}.
We chose to multiply the full LO cross section, including the Higgs and continuum diagrams, as well as their interference, with the $K$-factor computed for the signal process only (the $K$-factor calculations are described in detail below). There is an intrinsic uncertainty associated with this procedure, since the interference term receives higher order corrections at amplitude level that are different for the signal and the continuum background. 
This can possibly lead to a change in the relative phase of the interference term. While the sign of the latter can be judged, its size gathers some arbitrariness in the absence of a complete higher order computation of the continuum background. The uncertainty on the interference term associated to our procedure is estimated to be up to $30\%$ \cite{Bonvini:2013jha,Ellis:talk}.\footnote{We thank G.~Passarino and M~D\"uhrssen for comments about this point.}

We now describe the technical details of our simulations:

\begin{description}

\item {\bf Parton Distribution Function (PDF) sets and scales} ~The $gg\to ZZ$ process was simulated with LO PDFs. To reproduce the 8\,TeV result from CMS~\cite{CMS} we used the CTEQ6L set~\cite{Pumplin:2002vw} with factorization scale ($\mu_{\mathrm{fact}}$) and renormalization scale ($\mu_{\mathrm{ren}}$) equal to $m_{4\ell}/2$. As a consistency check, we verified that we reproduce the results of Ref.~\cite{Campbell:2013una}.\footnote{We performed the check with both MSTW2008 LO~\cite{Martin:2009iq} and CTEQ6L1 PDF, for the scale choices $\mu_{\mathrm{ren}}=\mu_{\mathrm{fact}}=m_{h}/2$ and $m_{4\ell}/2$.} 
The  rest of the results presented in this paper were obtained with the MSTW2008 LO PDF with scale choice $m_{4\ell}/2$. In all cases, the $q\bar{q}$-initiated background was simulated at NLO, using the NLO version of the same PDF used for the signal, and the same choice of scales. The acceptance cuts used in the CMS analysis~\cite{CMS} were adopted.

\item {\bf $K$-factors} ~Following the suggestion of Ref.~\cite{Bonvini:2013jha}, we applied to the $gg\to ZZ$ process the NNLO $K$-factor computed for inclusive production of a heavy SM Higgs. Specifically, we multiplied the LO cross section in each $m_{4\ell}$ bin with the NNLO $K$-factor computed for inclusive production of a SM Higgs with mass equal to the central value of the bin. The $K$-factors were obtained using the ggHiggs code~\cite{Ball:2013bra,ggHiggs}.\footnote{We used MSTW2008 (NN)LO for the (NN)LO cross sections, with scales set to $m_{h}/2$. We made use of the `finite-$m_{t}$' option available in the code. In the computation of the NNLO cross sections, all initial states were included up to NLO, and the $gg$ channel up to NNLO~\cite{Ball:2013bra}.} Table~\ref{tab:kfactor} lists the $K$-factors that were used for the different bins and different collider energies. Alternatively, and what would be a better prescription, one should use the $K$-factors computed for the invariant mass distribution of $gg\to h^{(*)} \to ZZ$ mediated by an off-shell $125$\,GeV Higgs, which can be somewhat different from those for inclusive production of a heavy Higgs~\cite{Passarino:2013bha}. However, by comparing with Ref.~\cite{Passarino:2013bha} we have checked that in the 8 TeV case the agreement is within $10\%$.   


Also notice that we made use of the $K$-factors computed for a heavy SM Higgs, even though the QCD corrections to the amplitudes proportional to $c_{t}$ and $c_{g}$ will be slightly different. As an estimate of this effect we computed the NLO $K$-factor for a heavy Higgs both for the measured value of the top mass and in the infinite top mass limit. We find that for a collider energy of 14\,TeV the $K$-factors differ by less than $10\%$, the one computed for $c_{g}$ being slightly larger.

\item {\bf Uncertainties} ~We wish to comment briefly on the theory uncertainties affecting our predictions for $gg\to ZZ$ at the 14\,TeV LHC. To estimate the scale uncertainties, we varied $\mu_{\mathrm{ren}}=\mu_{\mathrm{fact}}\in [m_{4_{\ell}}/4,m_{4\ell}]$, both in the LO cross sections and in the corresponding $K$-factors. The maximum variation of the cross section, over all the range of invariant masses considered in the analysis, is of $8\%$, which we take as our assessment of the scale uncertainty. As for PDF errors, we performed the following simple estimate: the $K$-factors were recomputed using two additional PDF sets (NNPDF2.3 NNLO~\cite{Ball:2012cx} and CT10 NNLO~\cite{Gao:2013xoa}) for the NNLO $pp\to h$ cross section, while keeping fixed the LO cross sections obtained with MSTW2008 LO. We found the maximum variation of the $K$-factors to be $\sim 5\%$, which we take as our estimate of the PDF uncertainty.\footnote{This estimate of the PDF errors applies also to all the FCC-hh energies we considered.} The scale and PDF uncertainties discussed here should be added to the intrinsic theory uncertainties related to the unknown exact higher order corrections to the $gg\to ZZ$ process, which were addressed above.


\end{description}

\begin{table}[t]
		\centering
		\begin{tabular}{|c|c|c|c|c|c|c|c|c|c|} \hline
	$\sqrt{s}$ [TeV]\,\,$\backslash$\, $m_{h}$ [GeV] & 325 & 500 & 700 & 950 & 1300 & 1750 & 2500 & 3500 & 4500 \\   
	\hline
	$14$ & 1.96 & 1.86 & 1.81 & 1.80 & 1.81 & $\ast$ & $\ast$ & $\ast$ & $\ast$ \\
	$33$ & 1.76 & 1.67 & 1.65 & 1.66 & 1.67 & 1.70 & 1.73 & 1.76 & 1.79 \\
	$50$ & 1.66 & 1.58 & 1.56 & 1.57 & 1.60 & 1.63 & 1.67 & 1.70 & 1.73 \\
	$80$ & 1.54 & 1.47 & 1.46 & 1.47 & 1.50 & 1.54 & 1.58 & 1.63 & 1.66 \\
	$100$ & 1.49 & 1.41 & 1.41 & 1.42 & 1.46 & 1.49 & 1.54 & 1.59 & 1.62 \\
	\hline
		\end{tabular}
		\caption{NNLO $K$-factors for inclusive production of a heavy SM Higgs that were used to rescale the LO $gg\to ZZ$ cross sections.}
	\label{tab:kfactor}
\end{table}

\noindent We would like to remark that a fully consistent computation of Higgs-mediated four-lepton production at $O(\alpha_{s}^2)$ would need to include the interference of the $qg$-initiated Higgs and continuum diagrams \cite{Campbell:2013una}. However, in Ref.~\cite{Campbell:2013una} this effect was found to be negligible in the high invariant mass range for a collider energy of 8\,TeV. Since we do not expect the relative size of the $qg$ channel to increase at higher collider energies, we neglected this effect in our analysis.

Recently, interesting progress has been made towards a computation of the two-loop contribution to the continuum amplitudes for both the interfering and non-interfering background \cite{Henn:2014lfa,Caola:2014lpa,Cascioli:2014yka}. In particular, in Refs.~\cite{Henn:2014lfa,Caola:2014lpa} both the planar and non-planar master integrals needed for the two-loop computation of $gg\to VV$ have been calculated, for massless fermions in the internal lines. While the massless approximation is certainly suitable for the light quarks, including the bottom, it is not appropriate for the top quark. In particular, we remark that at one-loop the contribution to the amplitude for $gg\to ZZ$ of the box diagrams with the quark $q$ running in the loop diverges at large $\hat{s}$ as $\sim (m_{q}^{2}/m_{Z}^{2})\log^{2}(\hat{s}/m_{q}^{2})$. 
This shows that, at least at one-loop, the top mass effects are relevant in the large-$\hat{s}$ region, on which our analysis is focused. A complete calculation of $gg\to ZZ$ at NLO, i.e. at $O(\alpha_{s}^{3})$, would require the computation of two-loop diagrams with a massive internal fermion, which is a challenging task with current technology. In any case, it is reasonable to expect further progress in the near future towards an NLO computation of the $gg\to ZZ$ interfering background. This is particularly important for the interference term, where the higher order corrections can possibly induce a shift in the relative phase.

 Because it is extremely difficult to guess the level of theoretical development, and therefore the level of accuracy of the predictions, that will be attained on time scales as long as those of the HL-LHC and FCC-hh, in the analysis of the upcoming sections we have ignored theoretical uncertainties.  However, in Section~\ref{sec:HLLHC} we compare the results with and without theoretical errors and find that with $3\, \hbox{ab}^{-1}$ at 14\,TeV the statistical errors are still dominant.

\subsection{Results for the HL-LHC} 
\label{sec:HLLHC}
Now we can proceed to the discussion of the precision of the 14\,TeV high-luminosity LHC. In order to thoroughly explore the different $\sqrt{\hat s}$ dependence of the contributions generated by $(c_t,c_g)$ we introduce the new binning for the four-lepton invariant mass 
\bea
\hbox{Binning} \ \ \sqrt{\hat s} =(250,400,600,800,1100,1500)\,\hbox{GeV}\, .
\eea
Then using the modified version of MCFM we calculate the event yields as functions of the $c_t,c_g$ parameters. The yields at $3$\,ab$^{-1}$ for the signal and the non-interfering background are reported in Eqs.~(\ref{yields14cpe},\,\ref{bcg})
\bea
\label{yields14cpe}
N [250,400]&=&{521 c_g c_t+187. c_g^2-491. c_g+381 c_t^2-687. c_t}+7044\,,\nonumber\\
N[400,600]&=&{394 c_g  c_t+143 c_g^2-229. c_g+423 c_t^2-564 c_t}+1136\,,\nonumber\\
N[600,800]&=&{97 c_g c_t+81 c_g^2-40 c_g+139 c_t^2-210c_t}+221\,,\\
N[800,1100]&=&{23. c_g c_t+65 c_g^2+3.6 c_g+59 c_t^2-100 c_t}+80\,,\nonumber\\
N[1100,1500]&=&{-2.4 c_g c_t+40. c_g^2+11.3 c_g+16.5   c_t^2-31 c_t}+22\,, \nonumber
\eea
\bea
\label{bcg}
N_{q\bar{q} \to ZZ}=(31410, 6904, 1417, 515, 145)\,.
\eea

\begin{figure}[t]
\begin{center}
\includegraphics[scale=0.65]{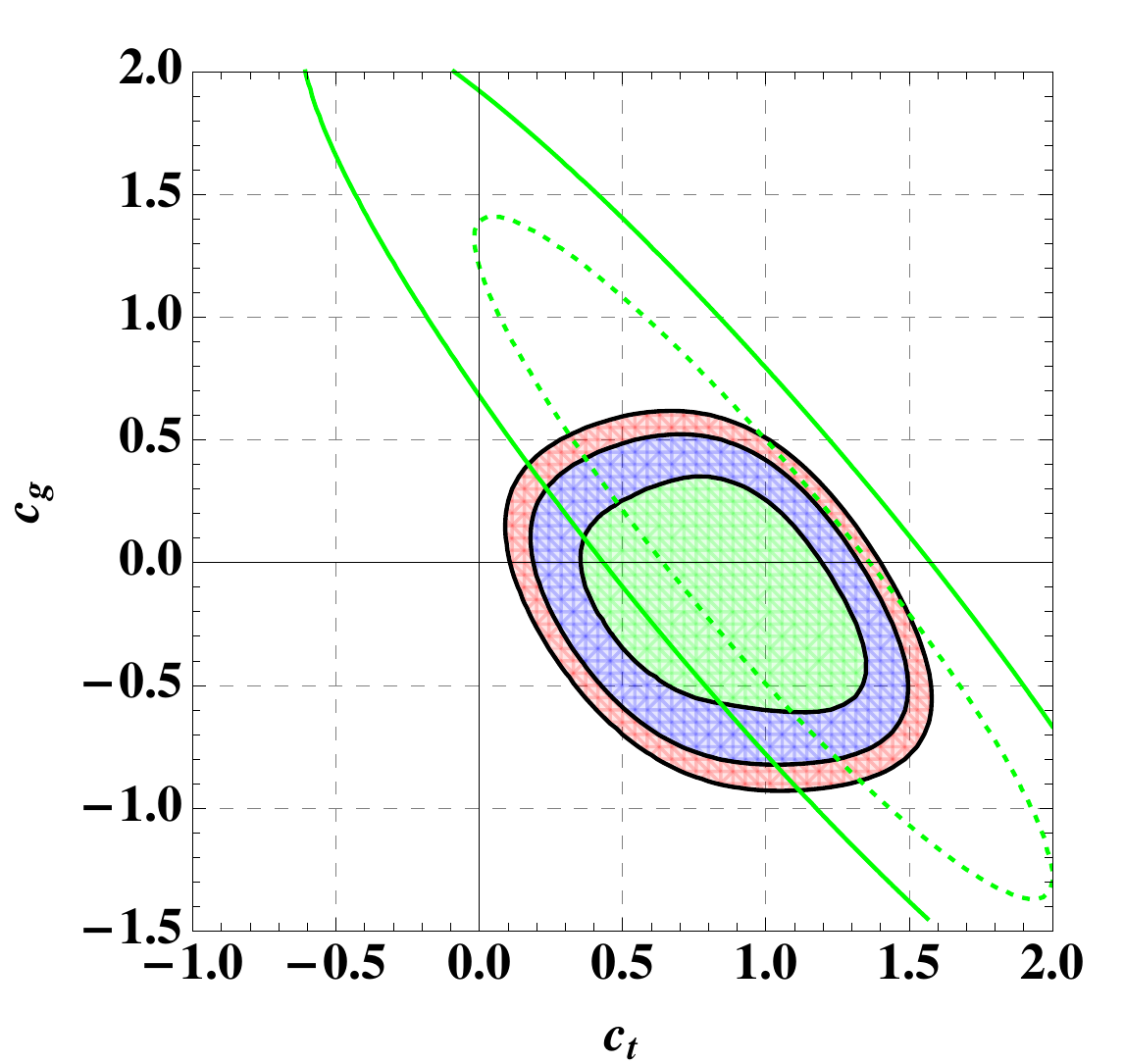}
\end{center}
\caption{\label{ctcg14} Prospects for a 14\,TeV analysis with an integrated luminosity of 3\,ab${}^{-1}$ and for the injected SM signal: $68\%,95\%$ and $99\%$ expected probability regions in the $(c_t,c_g)$ plane. The dashed and solid green lines indicate the $68\%$ and $95\%$ contours for the linear analysis, respectively. No theoretical uncertainty is included.}
\end{figure}

\begin{figure}[h]
\begin{center}
\includegraphics[scale=0.75]{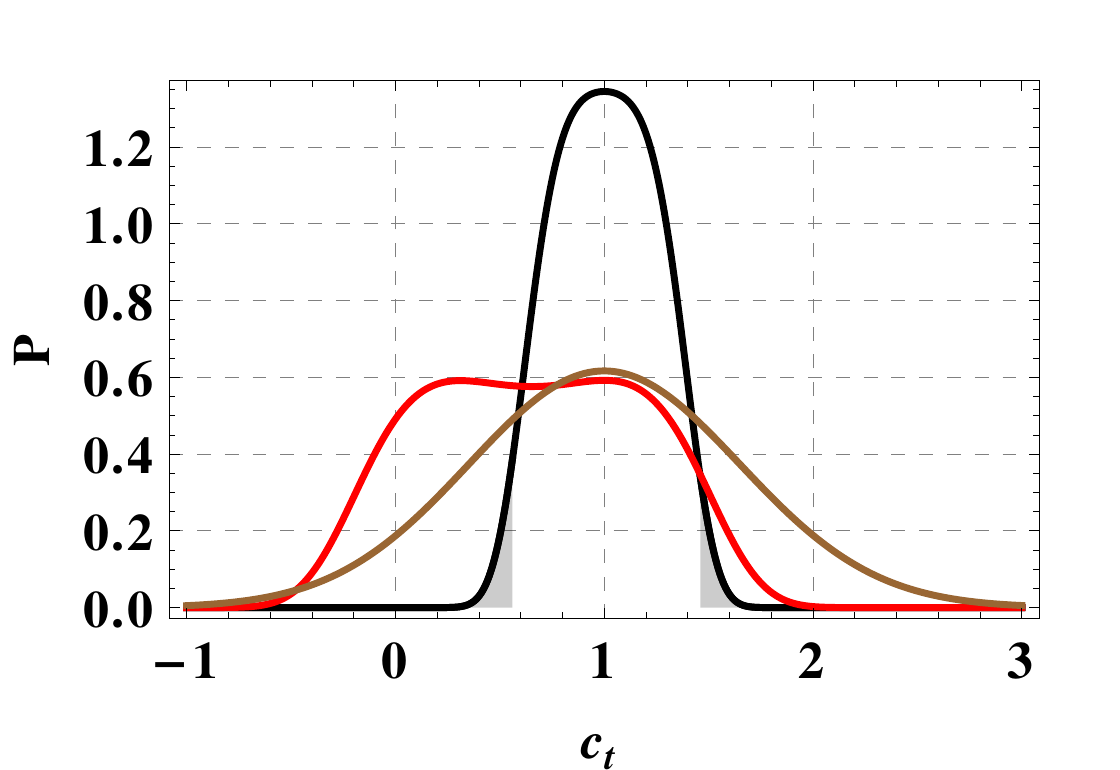}
\end{center}
\caption{\label{ct14} Prospects for the 14\,TeV analysis with an integrated luminosity of 3\,ab${}^{-1}$ and for the injected SM signal: expected posterior probability as a function of $c_t$, assuming the constraint $c_t+c_g=1$ and to observe the SM signal. The black curve corresponds to the nonlinear analysis including all bins, at 95$\%$ probability we find $c_t\in[0.56,1.46]$ (unshaded region), at 68$\%$ $c_t\in[0.74,1.28]$. The red curve was obtained using only the categories below $600$\,GeV and at $68\%$ we have $c_t\in[0.1,1.25]$. The brown curve corresponds to the linear analysis including all bins, which gives $c_t\in[0.36,1.66]$ at $68\%$.}
\end{figure}

The corresponding probability contours are reported in Fig.~\ref{ctcg14}, for both the nonlinear and linear analyses. Differently from the 8\,TeV case, at the HL-LHC the EFT treatment is meaningful, since the nonlinear analysis is powerful enough to constrain the Wilson coefficients to be $<1$. However, as it was discussed in Section~\ref{sec:prod_ops}, the validity of the nonlinear analysis depends on the relative size of the dimension-6 and dimension-8 coefficients, see Eq.~\eqref{lvsnl}, and as such the nonlinear results are still model-dependent.  We will discuss in Section~\ref{sec:toymodel} one example model where the nonlinear analysis does apply. The linear bounds, which are truly model-independent, are significantly weaker. To make explicit the $c_t$ vs.~$c_g$ differentiating power of our analysis we have also studied the one-dimensional probability obtained by fixing $c_t+c_g=1$, the results are presented in Fig.~\ref{ct14}. We can see that with our simplistic analysis we can constrain $c_t$ to be within $[0.75,1.28]\,([0.56,1.46])$ with $68\%\,(95\%)$ probability. This result was derived using the nonlinear analysis, whereas in the linear approach we find $c_t\in[0.36,1.66]$ with $68\%$ probability. 
 The results presented above were obtained assuming zero systematic uncertainty. Assuming $30\%$ theoretical error on the total $gg\ra ZZ$ cross section the bound on $c_t$ is relaxed to $[0.74,1.3]$ with $68\% $ probability.
 One can see that our counting analysis is dominated by the statistical error, however the theoretical uncertainties will become a serious limitation once we move to higher precision, either by implementing the MELA analysis or by studying the prospects of the future colliders.

Lastly, we observe that at larger luminosity  $\gtrsim 30$\,ab$^{-1}$ the differences between the linear and nonlinear analysis are reduced, their respective bounds on $c_{t}$ differing by less than $20\%$.

\subsection{Bounds on top partners} \label{sec:toymodel}

\begin{figure}
\begin{center}
\includegraphics[scale=0.7]{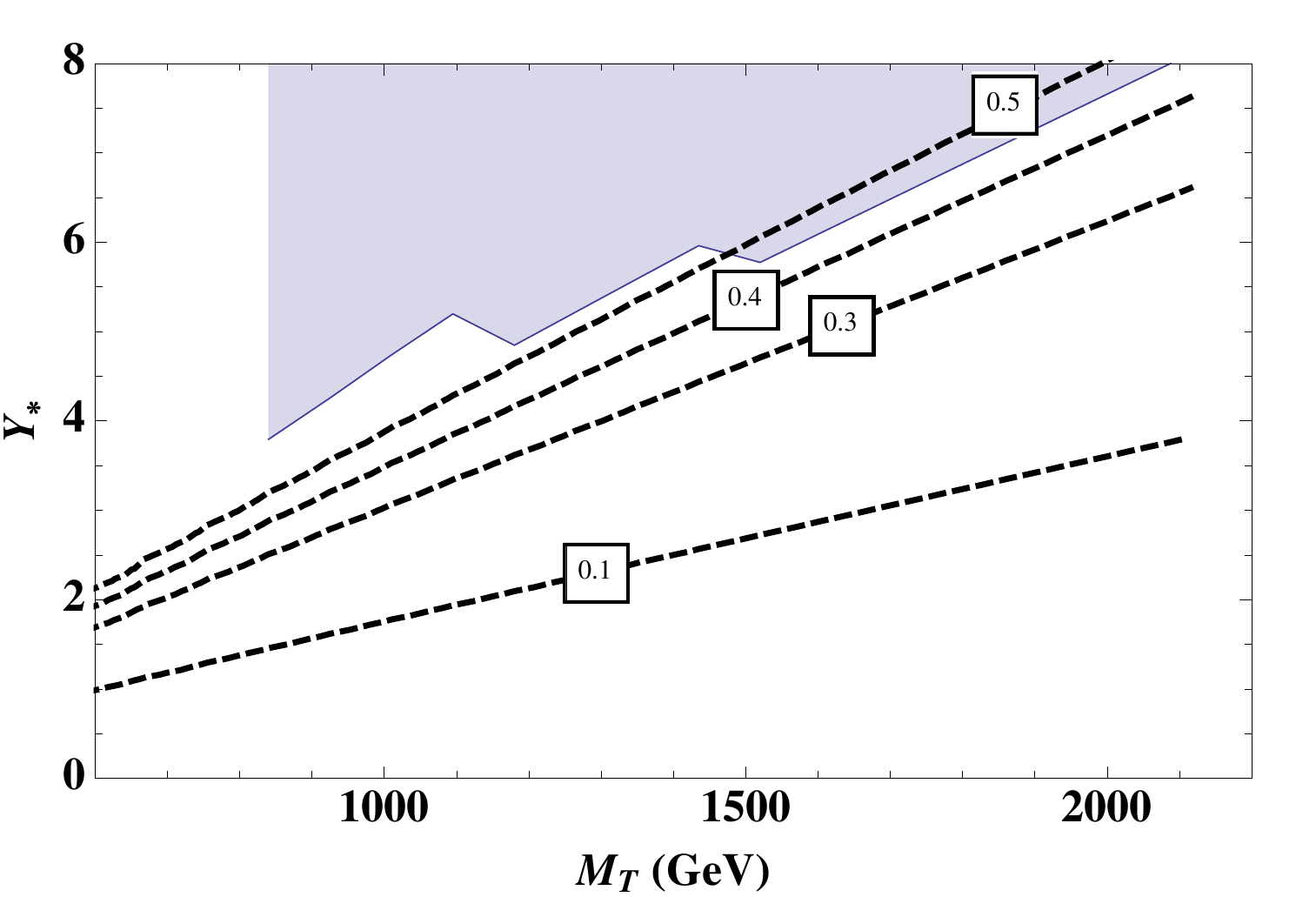}
\caption{
\label{YeM}The shaded region shows the $95\%$ expected exclusion in the top partner parameter space at the HL-LHC. $M_T$ denotes the physical mass of the top partner. The black dashed lines indicate the isocontours of $c_g$.}
\end{center}
\end{figure}

%
The $c_t$ vs.~$c_g$ degeneracy arises in models with fermionic top partners, in particular it is generic in the composite Higgs models~\cite{Falkowski:2007hz,Low:2010mr,Azatov:2011qy ,Delaunay:2013iia,Montull:2013mla}. As a prototype of the models with this degeneracy we can introduce just one vector-like top partner $T$, transforming as a singlet of $SU(2)_{L}$
\bea
\label{tprime}
-{\cal L}\,=\,y\bar Q_L \widetilde{H} t_R + Y_*\bar Q_L \widetilde{H} T_R +M_*\bar T_L T_R + \mathrm{h.c.}\,.
\eea
In this model, loops of the heavy fermion $T$ generate an effective interaction of the Higgs with the gluons, and at the same time the top Yukawa coupling is modified due to the mixing with the top partner. Due to the Higgs low-energy theorem, the on-shell Higgs production cross section is predicted to be the same as in the SM, since it can easily be checked~\cite{Azatov:2011qy,Delaunay:2013iia} that, after integrating out the heavy top partner, $c_{t}+c_{g}=1$. 
%
%
%
Besides modifying the Higgs-mediated amplitude for $gg\to ZZ$, the $T$ also enters in the box diagrams, generating a contribution to the interfering background which in the EFT must be parameterized by a dimension-8 operator. We can estimate the Wilson coefficients of the dimension-6 and dimension-8 operators in Eqs.~\eqref{d6} and \eqref{d8} as 
\bea
&c_g=c_y\sim \frac{Y_*^2 v^2}{M_*^2}\,,\nonumber\\
&c_8\sim \frac{Y_*^2 v^4}{M_*^4}\,.
\eea
This implies that the dimension-8 operators will become important at the scale
\bea
\sqrt{s}\sim M_*\,,
\eea
where our analysis breaks down.\footnote{
 As a side comment, we note that an exact treatment of the $gg\to ZZ$ amplitude in this model requires the computation of box diagrams with two different massive fermions in the loop. These diagrams are exactly the same as those for the SM contribution to the $gg\to WW$ process, mediated by top and bottom quarks \cite{Duhrssen:2005bz}. Within this work, however, we chose to remain within the EFT approach and leave the analysis of the effects of the dimension-8 operators for future study.
} 
Therefore to remain in the region of validity of the EFT approach, when deriving the bounds on the model parameter space we only considered the bins with invariant mass below the physical mass of the top partner, $M_{T}$. Since the model depends only on two free parameters once the top mass is fixed, we can plot the exclusion contours in the $(Y_*, M_T)$ plane. The result, obtained applying the nonlinear analysis, is shown in Fig.~\ref{YeM}. As can be seen from the figure, the bound applies to a region with large Yukawa coupling, $Y_{\ast}\gg 1$: this implies that $c_{g,y}^{2}\gg c_{8}$, thus justifying the use of the nonlinear analysis. We note that the simple model described by Eq.~\eqref{tprime} is equivalent (as far as the $gg\ra h^{(*)} \ra ZZ$ process is concerned) to the recently proposed simplified composite Higgs models ${\bf M1_{5,14}}$
\footnote{The composite Higgs models mentioned here are based on the $SO(5)/SO(4)$ coset. The label ${\bf 1,4}$ indicates the $SO(4)$ representation in which the top partners transform, while the underscript ${\bf 5,14}$ specifies the representation of $SO(5)$ in which the $Q_{L}$ doublet is embedded.}
 of Ref.~\cite{DeSimone:2012fs}, in the limit $v\ll f$. Similar bounds on the models ${\bf M4_{5,14}}$~\cite{DeSimone:2012fs} appear to be irrelevant, since in these scenarios the masses of the top partners are correlated with their Yukawa couplings and large values of the Yukawa couplings can appear only at the price of increasing the heavy fermion masses.

\subsection{$CP$-odd case}

So far we have been focusing on the $CP$-even operators. Let us now turn our attention to the $CP$-odd operators: the $CP$-odd Lagrangian after the electroweak symmetry breaking becomes%
\bea
{\cal L}&=&i \tilde c_t \frac{m_t}{v}\bar t \gamma_5 t h+\tilde c_g \frac{g_s^2}{32\pi^2}G_{\mu\nu}^a\tilde G_{\mu\nu}^a\,,\nonumber\\
\tilde c_t&=&\mathrm{Im}(c_y)\,.
\eea
Since the new physics contribution is $CP$-odd, it does not interfere either with the Higgs-mediated or with the continuum $gg\ra ZZ$ SM amplitudes. Rather than implementing the $CP$-odd operators in MCFM, we made the assumptions that the acceptance and $K$-factors are the same as in the $CP$-even case, and simply rescaled the $CP$-even results using the expressions of the LO matrix elements (see Appendix \ref{sec:loop} for the loop functions). The yields at $3$\,ab$^{-1}$ as functions of $\tilde c_t,\tilde c_g$ are reported in Eq.~\eqref{cpodd}
\bea
\label{cpodd}
N[250,400]&=&{1442 \tilde{c}_g \tilde{c_t}+434 \tilde{c}_g^2+1383 \tilde{c_t}{}^2 }+6740\,,\nonumber\\
N[400,600]&=&{598 \tilde{c}_g \tilde{c_t}+326 \tilde{c}_g^2+905\tilde{c_t}{}^2}+996\,,\nonumber\\
N[600,800]&=&{73\tilde{c}_g \tilde{c_t}+181 \tilde{c}_g^2+207\tilde{c_t}{}^2}+150\,,\\
N[800,1100]&=&{-7.49\tilde{c}_g \tilde{c_t}+146 \tilde{c}_g^2+78 \tilde{c_t}{}^2}+39\,,\nonumber\\
N[1100,1500]&=&{-18.2 \tilde{c}_g \tilde{c_t}+88 \tilde{c}_g^2+20.\tilde{c_t}{}^2}+7.6\,. \nonumber
\eea
\begin{figure}
\begin{center}
\includegraphics[scale=0.7]{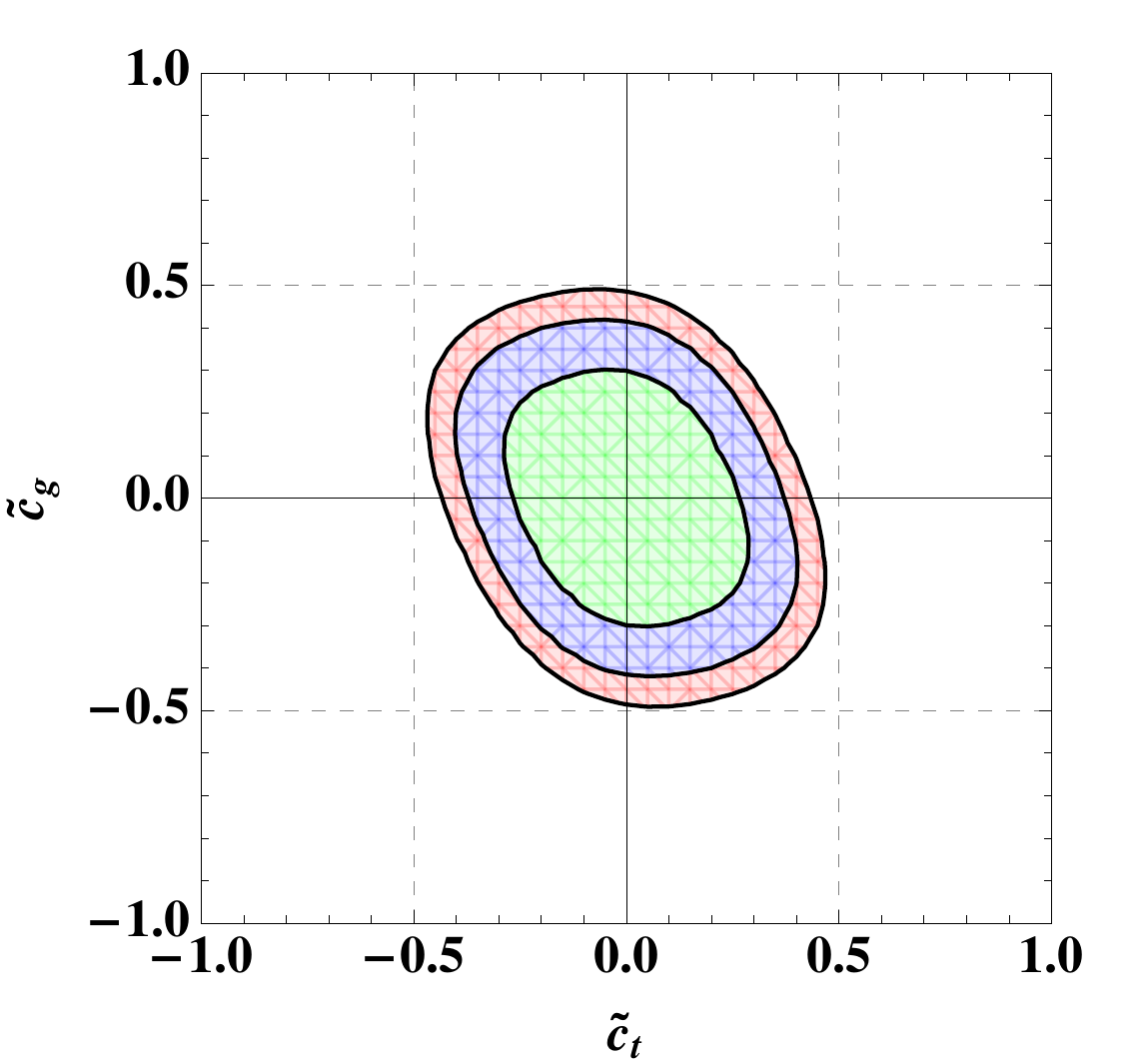}
\caption{\label{cpo} Prospects for a 14\,TeV analysis with an integrated luminosity of 3\,ab${}^{-1}$ for the injected SM signal: $68\%,95\%$ and $99\%$ expected probability contours in the $(\tilde c_t,\tilde c_g)$ plane.  
}
\end{center}
\end{figure}
The constraints in the $(\tilde c_t, \tilde c_g)$ plane are presented in Fig.~\ref{cpo}. This analysis is valid under the assumption that dimension-6 and dimension-8 $CP$-even effects are subleading with respect to the $CP$-odd contributions considered here, and thus the derived bounds are not truly model-independent.

\subsection{Results for the FCC-hh}

Finally we would like to comment on the prospects of the FCC-hh on the studies of the $c_t,c_g$ couplings. We present our estimates for $33,50,80$ and $100$\,TeV proton-proton colliders, assuming an integrated luminosity of $3$\,ab$^{-1}$. In our analysis we have used exactly the same acceptance cuts as for the 8 and 14\,TeV LHC. This assumption is quite likely to be unrealistic, nevertheless our results can be considered as first estimates of the range which can be tested at the future high-energy proton-proton colliders. To perform this analysis we modified the binning to
\bea
\hbox{Binning}\ \ \sqrt{\hat s}=(250,400,600,800,1100,1500,2000,3000,4000,5000)\,\mathrm{GeV}\, .
\eea
The results of our analysis are presented in Table~\ref{tabfcc}, under the assumption that $c_{t}+c_{g}=1$. We can see that as we go to higher collider energy the differences between linear and nonlinear probabilities decrease, and strong model-independent bounds on $c_{t}$ are obtained. 
\begin{table}
\begin{center}
\begin{tabular}{ |c|c|c|c|c|}
\hline
   & 33\,TeV & 50\,TeV & 80\,TeV & 100\,TeV \\
 \hline
   non-linear $<2$TeV &[0.92,1.14] & [0.95,1.11] & [0.96,1.08]&[0.97,1.07] \\
  \hline
 linear $<2$\,TeV& [0.83,1.18] & [0.9,1.11] & [0.94,1.07] &[0.95,1.05]\\
 \hline
   non-linear all &[0.94,1.11] & [0.96,1.08] & [0.98,1.05]&[0.98,1.04] \\
  \hline
 linear all& [0.84,1.16] & [0.91,1.09] & [0.95,1.05] &[0.96,1.04]\\
 \hline
\end{tabular}
\caption{$68\%$ probability intervals on the value of $c_t$, obtained assuming $c_t+c_g=1$ and injecting the SM signal at various collider energies. In all cases an integrated luminosity of 3\,ab${}^{-1}$ was assumed. The numbers in the second and the third row present the non-linear and linear analysis, respectively, for the low-energy bins only, $\sqrt{\hat{s}}<2$\,TeV. The fourth and the fifth rows contain the corresponding numbers obtained including all the bins up to $5$\,TeV.
\label{tabfcc}}
\end{center}
\end{table}

\section{Conclusion}
\label{sec:conclusion}
%
We wish to summarize briefly the main results of our paper. We have discussed the implications of the $pp\ra h^{(*)} \ra Z^{(*)} Z^{(*)}\ra 4 \ell$ measurement at high center-of-mass energy on the Wilson coefficients of the dimension-6 operators modifying the Higgs interactions.  We have shown that this process is especially powerful in constraining the two dimension-6 operators contributing to the Higgs production in gluon fusion, which parameterize the modifications of the top Yukawa coupling and the effective interactions between the Higgs boson and the gluons mediated by heavy new physics. The sum of these two effects is already constrained by inclusive Higgs measurements, whose agreement with the SM implies the approximate relation $c_{t}+c_{g}\sim 1$. However, the current bounds on each of the two operators individually are very weak, because the precision is controlled by the $pp\ra t\bar {t} h$ process, where $O(1)$ deviations are still allowed. The recent measurement by CMS of $pp\ra Z^{(*)} Z^{(*)}\ra 4 \ell$ at large invariant mass of the four leptons, which receives contributions from off-shell Higgs exchange, provides us with a new way to measure  the Higgs effective interactions. Combining on-shell and off-shell data should thus make it possible to disentangle the effects of $c_{t}$ and $c_{g}$. 

Wherever applicable, we have discussed our results in the EFT language, rather than in terms of a simple anomalous couplings parameterization. In particular, we have derived the conditions under which the dimension-8 operators can be safely ignored, which allowed us to understand the range of validity of our results. This type of self-consistency check comes as a bonus of the EFT approach.


We have obtained the first constraints on the modifications of the top Yukawa coupling, $c_{t}$, by recasting the CMS 8\,TeV bound on the Higgs width~\cite{CMS}. These constraints are weaker than those currently available from the direct $t\bar{t}h$ measurement, but roughly of the same order. Since $O(1)$ corrections to the SM are still allowed, no EFT interpretation is possible with current data. 

Next, the possibilities of the HL-LHC in measuring $c_{t},c_{g}$ were explored. We have found that at $14$\,TeV collision energy and 3\,ab${}^{-1}$ luminosity it will be possible to measure $c_{t}$ with $\sim 25\%$ precision. Even though this estimate is worse than the current prospects on the top Yukawa coupling precision measurements~\cite{ATLAS-Collaboration:2012jwa} from $t\bar{t}h$, we would like to stress that the off-shell measurements test
  roughly the same region of the  parameter space, and that there is still significant room for improvements by performing the dedicated matrix element analysis, which exploits all the angular information available in the four-lepton final state. As a caveat, we found that the 14\,TeV bounds can be altered by the presence of dimension-8 operators, if the new physics is weakly coupled. We have also presented the HL-LHC exclusion prospects for a toy prototype of the widely-studied Composite Higgs models, as well as constraints on the $CP$-violating Higgs interactions. 

Along the way, we addressed the status of current theoretical predictions for the $gg\to ZZ$ process, which suffer primarily from the lack of a computation of higher-order QCD corrections to the box diagrams. We described our choice of the procedure for approximating these corrections, which consists in applying the $K$-factor computed for the Higgs-mediated diagrams to the entire $gg\to ZZ$ cross section. 


Lastly, we commented on the reach of the future proton-proton colliders with energies between $33$ and $100$\,TeV. There a measurement of $c_{t}$ to $\sim 5\%$ accuracy is possible already within our simplistic study 
(assuming zero theoretical uncertainty), 
and the EFT analysis shows that the bounds obtained are fully model-independent.

\vspace{.2cm}
\noindent {\bf Note added} ~While this work was being completed, an independent study appeared~\cite{Cacciapaglia:2014rla} which also proposed to use the Higgs off-shell data to break the $c_t,c_g$ degeneracy. 

\vspace{4mm}

\noindent{\bf Acknowledgments} ~We thank M.~Bonvini for very useful discussions and for expanding the functionality of ggHiggs to accommodate our requirements. We would also like to thank V.~del Duca, G.~Passarino and L.~Reina for useful discussions and M.~D\"uhrssen and R.~Contino for comments about the manuscript. A.P. would like to acknowledge the University of Notre Dame du Lac, and specially D.~Patel, for providing computational resources.  A.P. is funded by the European Research Council under the European Union's Seventh Framework Programme (FP/2007-2013)/ERC Grant Agreement n.~279972. E.S. was supported in part by the US Department of Energy under grant DE-FG02-91ER40674, and wishes to thank the Institute for Theoretical Physics of the U.~of Heidelberg for hospitality during part of this project. A.A. thanks the  U.~of Rome ``La Sapienza'' for hospitality during part of this project. C.G. is  supported by the European Commission through the ERC Advanced Grant 226371 MassTeV and the Marie Curie Career Integration Grant 631962, by the Spanish Ministry MICINN under contract FPA2010--17747 and by the Generalitat de Catalunya grant 2014--SGR--1450. A.A. and C.G. thank the Centro de Ciencias de Benasque Pedro Pascual for its hospitality while part of this work was being carried out.

\section*{Appendices}

\appendix
\section{Fitting the Higgs width}

\label{cmsHwidth}
In this section we shall derive the bound on the Higgs width using the CMS data. The difference between our  result and the official analysis can be a measure of accurateness of our method. We perform the analysis based only on the counting experiment data presented in Ref.~\cite{CMS}, Fig.~1(a). The off-peak event yield is proportional to
\bea
\label{width}
&&{N_{\hbox{\small off peak}}}\sim g^4  A+ g^2 B+C,\nonumber
\eea
where $g$ stands for a universal rescaling of the SM couplings. 
The coefficients $A,B,C$ are related to the functions $F_i$ in Eq.~\eqref{Fi} in the following way
\bea
A&=&\int d m_{4 l} [F_1(m_{4\ell})+F_2(m_{4\ell})],\nonumber\\
B&=&\int d m_{4 l} [F_3(m_{4\ell})+F_4(m_{4\ell})],\\
C&=&\int d m_{4 l} F_0(m_{4\ell})\,.\nonumber
\eea
The requirement of keeping the number of on-peak events fixed to the SM value leads to the constraint $g^4/\Gamma = \mathrm{constant}$, so that we can parametrize the off-peak event yield as
\bea
{N_{\hbox{\small off peak}}} =A\, \frac{\Gamma} {\Gamma_{SM}}+ B\sqrt{\frac{\Gamma}{\Gamma_{SM}}} + C\,.
\eea
To calculate the functions $F_i$ we use MCFM and the same PDF set adopted by CMS, namely the CTEQ6L. We digitize Fig.~1(a) from Ref.~\cite{CMS} to extract the $q\bar{q}\ra ZZ$ background, as well as the observed number of events (see Table~\ref{dataT}). Following the prescription by CMS we apply a $m_{4\ell}$-independent $K$-factor of $2.7$ to the signal and the interfering background. We assume an average acceptance of $95\%$ for each lepton and with these numbers we are able to reproduce the reported CMS yields within $\sim 10 \% $. Then we perform a Bayesian analysis with 10$\%$ systematic uncertainty on the noninterfering background, which leads to the following bound on the Higgs width
\bea
\Gamma< 24.6\, \Gamma_{\mathrm{SM}}\,,
\eea
to be compared with the result quoted by CMS 
\bea
\Gamma<26.3 \,\Gamma_{\mathrm{SM}}.
\eea
For the sake of completeness, we present the event yields as functions of the $(c_t,c_g)$ couplings for the 8\,TeV analysis:
\bea
N_{[220,240]}&=&0.19c_g c_t+0.09 c_g^2-0.42 c_g+0.11 c_t^2-0.47 c_t+8.68\,,
 \nonumber\\
N_{[220,265]} &=&0.22 c_g c_t+0.10c_g^2-0.37c_g+0.13 c_t^2-0.43 c_t+7.38\,,\nonumber\\
N_{[265,295]}&=&0.24 c_g c_t+0.10 c_g^2-0.30c_g+0.15 c_t^2-0.36c_t+5.34\,,\nonumber\\
N_{[295,330]}&=&0.26 c_g c_t+0.10 c_g^2-0.24 c_g+0.17c_t^2-0.31 c_t+3.52\,,\nonumber\\
N_{[330,370]}&=&0.30c_g c_t+0.10 c_g^2-0.22c_g+0.24c_t^2-0.34 c_t+2.19\,,\nonumber\\
N_{[370,410]}&=&0.28 c_g c_t+0.08 c_g^2-0.18 c_g+0.26 c_t^2-0.34 c_t+1.25\,,\nonumber\\
N_{[410,460]}&=&0.27 c_g c_t+0.08c_g^2-0.16 c_g+0.27c_t^2-0.35 c_t+0.90\,,\\
N_{[460,520]}&=&0.21c_g c_t+0.08c_g^2-0.12 c_g+0.23c_t^2-0.31  c_t+0.58\,,\nonumber\\
N_{[520,580]}&=&0.13 c_g c_t+0.06c_g^2-0.07c_g+0.16 c_t^2-0.21   c_t+0.32\,,\nonumber\\
N_{[580,645]}&=&0.08c_g c_t+0.05c_g^2-0.04c_g+0.11 c_t^2-0.16 c_t+0.19\,,\nonumber\\
N_{[645,715]}&=&0.05c_g c_t+0.04 c_g^2-0.02 c_g+0.07 c_t^2-0.11 c_t+0.12\,,\nonumber\\
N_{[715,800]}&=&0.03 c_g c_t+0.04 c_g^2-0.01c_g+0.05c_t^2-0.08   c_t+0.08\,,\nonumber\\
N_{>800}&=&0.02c_g c_t+0.03 c_g^2-0.002c_g+0.03 c_t^2-0.06   c_t+0.05\,.\nonumber
\eea

\begin{table}
\begin{tabular}{|c|c|c|c|c|c|c|c|}
\hline
$m_{4\ell} \in$ [GeV]  & $\Gamma=\Gamma_{\mathrm{SM}}$&  $q\bar{q}$ bkg & data & $\Gamma=\Gamma_{\mathrm{SM}}$, reconstructed & $A$ & $B$ & $C$\\
\hline
[220,240]&8.4 &38.5&45&8.31 &0.11&$-0.47$&8.68\\
\hline
[240,265]&7.2&33.7&36&7.07&0.13&$-0.44$&7.38\\
\hline
[265,295] &5.4&27&31&5.12&0.15&$-0.36$&5.33\\
\hline
[295,330]&3.6&20&17&3.39&0.18&$-0.31$&3.52\\
\hline
[330,370]&2.2&13.9&16&2.08&0.24&$-0.35$&2.19\\
\hline
[370,410]&1.2&9.6&9&1.17&0.26&$-0.34$&1.25\\
\hline
[410,460]&0.9&6.2&11&0.81&0.27&$-0.35$&0.90\\
\hline
[460,520]&0.6&4.1&6&0.51&0.23&$-0.31$&0.58\\
\hline
[520,580]&0.3&2.6&6&0.26&0.16&$-0.21$&0.32\\
\hline
[580,645]&0.2&1.7&3&0.15&0.11&$-0.16$&0.19\\
\hline
[645,715]&0.1&1.1&2&0.08&0.07&$-0.11$&0.12\\
\hline
[715,800]&0.09&0.7&1&0.05&0.05&$-0.08$&0.08\\
\hline
$>$800&0.2&1&0&0.03&0.03&$-0.06$&0.05\\
\hline
\end{tabular}
\caption{\label{dataT}The digitized data from Ref.~\cite{CMS} as well as the values of the coefficients $A,B,C$ reconstructed using MCFM. The columns from the second to the fourth one contain the results of the digitization of Fig.~1(a) in Ref.~\cite{CMS}. The fifth to eight columns show the results of our MCFM simulations. The fifth column is the reconstructed yield for $gg\ra ZZ$, which we present as a cross check against the CMS numbers. Note that the precision of our digitization is $\sim 0.2$ for the number of events and is limited by the resolution of the plot.}
\end{table}
%

\section{Operators modifying the Higgs decay}  
\label{hdecay}

We now want to examine the operators that would modify the Higgs couplings to the $Z$ bosons. The off-shell measurements can constrain  more effectively the operators which grow with energy. Let us consider  the following operator
\bea
\label{cBox}
O_{\Box}=\frac{c_{\Box}}{v}\Box h Z_{\mu} Z^{\mu}\,. 
\eea
Then the signal rate will be modified as (keeping only the terms linear in $c_\Box$)
\bea
N_{\hbox{\small off-peak}} \simeq A\l(1-2c_\Box\frac{m_{4\ell}^2}{m_Z^2}\r)+B\l(1-c_\Box\frac{m_{4\ell}^2}{m_Z^2}\r) + C\, ,
\eea
where the coefficients $A,B,C$ were defined in Appendix~\ref{cmsHwidth}. Then we find 
\bea
&&68 \%: ~c_{\Box}\in[-0.7,-0.17]\cup[0.42,0.84]\,,\nonumber\\
&&95\%:~c_{\Box}\in [-0.96,0]\cup[0.21,1.15]\,.
\eea
However, if the Higgs boson is part of an $SU(2)_{L}$ doublet, the operator~(\ref{cBox}) can originate only from the following gauge-invariant dimension-8 operator:
\bea
\frac{(D_\mu H)^2 \Box(H^\dagger H)}{\Lambda^4}\,,
\eea
therefore the bounds on the scale will be irrelevant, $\Lambda \gtrsim 150$\,GeV. At the dimension-6 level there are the following operators modifying the Higgs interactions with the $Z$ boson
\bea
\l(D_\mu H\r)^\dagger \sigma^a D_\nu H W^{\mu\nu, a},~~\l(D_\mu H\r)^\dagger D_\nu H B^{\mu\nu}, ~~H^\dagger H B_{\mu\nu} B^{\mu\nu}\, ,\nonumber\\
\l(H^\dagger \sigma^a \dblarrow{D}_\nu H\r) (D^\mu W_{\mu\nu})^a,~~\l(H^\dagger \dblarrow{D}_\nu H\r) (D^\mu B_{\mu\nu}),
\eea
which lead to the interactions 
\bea
h Z_{\mu\nu}Z^{\mu\nu},~~h Z_{\mu} \d^\nu Z^{\mu\nu}\, .
\eea
However none of these operators  will affect the longitudinal components of the $Z$, therefore the overall growth of the amplitude with the energy is the same as in the SM. As a consequence, going to high energy does not lead to a strong enhancement of the signal.

\section{Loop functions}
\label{sec:loop}
For the sake of completeness we report the $CP$-even and $CP$-odd loop functions for the triangle diagrams \cite{Ellis:1975ap,ShifmanETAL,Djouadi:2005gi}. The $CP$-even $F_{\Delta}$  and $CP$-odd $\tilde F_{\Delta}$ loop functions are given by
\bea
&&F_\Delta(m)=\frac{3}{2\tau^2}[\tau+(\tau-1)f(\tau)],\qquad \tilde F_{\Delta}(m)=\frac{f(\tau)}{\tau},\nonumber\\
&&\tau=\frac{\hat{s}}{4m^2},\quad 
f(\tau)=\l\{\baa{c}\arcsin^2 \sqrt{\tau}\ \ \hbox{for }~\tau \leq 1, \\
-\frac{1}{4}\l[\log\frac{1+\sqrt{1-1/\tau}}{1-\sqrt{1-1/\tau}}-i \pi\r]^2\ \hbox{for}~\tau>1\,.\eaa\r.
\eea

\end{document}